# Research Intelligence (CRIS) and the Cloud: A Review

Otmane Azeroual[1] , & Joachim Schöpfel[2]

[1] German Center for Higher Education Research and Science Studies (DZHW), Berlin, Germany

[2] GERiiCO Laboratory, University of Lille, Villeneuve-D'Ascq, France

Correspondence: Otmane Azeroual, German Center for Higher Education Research and Science Studies (DZHW), Schützenstraße 6a, 10117 Berlin, Germany. Tel: +49 302064177-99. E-mail: azeroual@dzhw.eu

**Abstract**

The purpose of this paper is to explore the impact of the cloud technology on current research information systems (CRIS). Based on an overview of published literature and on empirical evidence from surveys, the paper presents main characteristics, delivery models, service levels and general benefits of cloud computing. The second part assesses how the cloud computing challenges the research information management, from three angles: networking, specific benefits, and the ingestion of data in the cloud. The third part describes three aspects of the implementation of current research systems in the clouds, i.e. service models, requirements and potential risks and barriers. The paper concludes with some perspectives for future work. The paper is written for CRIS administrators and users, in order to improve research information management and to contribute to future development and implementation of these systems, but also for scholars and students who want to have detailed knowledge on this topic.

## 1. Introduction

Currently, the activities of universities, researchers, the results of their work are widely disseminated on the Internet. Specialized and transdisciplinary databases such as Scopus or Web of Science allow the exploitation and processing of these results. Locally and for research institutes, the management of their activity involves the use of a Current Research Information System (CRIS) or Research Information System (RIS). CRIS as Research Intelligence is a central database or federated information system that can be used to collect, manage and provide information on research activities and research results. The relevance of such a system is to be able to process a large volume and variety of data, to record thousands of daily interactions and to make relevant segmentation based on factual data, like projects, publications, patents, awards etc. (Azeroual et al. 2018; Azeroual & Schöpfel, 2019).

It is powered by different data sources from users. The latter after authentication, have access to the system via several interfaces and can enter their data or constantly update the database. It will be solicited by complex decision requests.

The sustainability of information systems is a permanent challenge that must be addressed, because of the risk of inherent and intrinsic problems (Leymann, 2009). In this context, data security appears to be a vital demand because of the challenges it presents and the issues it implies. This security request materializes on several operative levels of the treatment and circulation of information (Mell & Grance, 2011).

In the case of CRIS, university administrations and researchers attach great importance to the quality and security of their data (Azeroual & Abuosba, 2017). They also need a high speed of processing and a lot of storage space. Cloud computing appears as a possible solution to this problem.

Indeed, relieving research institutions from the responsibility of maintenance and back-up, enabling data sharing, offering the computing power of cloud servers and reducing the cost of implementing CRIS, constitute the major assets and arguments in favor of cloud computing as an option for research information management. On the other side, the security will remain a major issue because of the external hosting of important, strategic and partly sensitive data in the cloud, which often means outside of the research institute.

Today, more and more universities and research organizations are running a research information management system, or intend to do so in the future. Cloud computing is one of the options. Cloud computing has become a recurrent topic of papers and posters presented at the meetings and conferences on research information systems, from both system providers and system users. Based on this literature, and based also on own empirical work, the paper will provide an overview on the relevance of cloud computing for research information management.

After a review on literature on general characteristics, delivery models, service levels and potential benefits of cloud computing, the paper will assess three dimensions of the relationship between CRIS and the cloud: the cloud as a natural option for academic networking and research infrastructures (including the European research cloud), the cloud as a technological, operational and commercial alternative for CRIS, and the fact that more and more research information are available externally, hosted somewhere in the cloud.

The third part will describe service models, technical and other requirements and potential risks of implementing or opting for a research information system in the cloud.

In order to maintain a high level of performance in a competitive environment, research information systems will continue to change, they will develop and implement new approaches and methods of integration, modeling, querying databases. Cloud computing is already part of the equation. We are convinced that in the future it will become even more important for research information management.

**2. Cloud Computing**

Adopted and popularized by large corporate companies like Google, Amazon and Microsoft about 20 years ago, cloud computing has become common practice in today's computer world. The term refers to "*both the applications delivered as services over the Internet and the hardware and systems software in the data centers that provide those services*" (Fox et al. 2009, pp. 4). The concept of the cloud has been described as "*a simple one: a service provider processes, manages or stores customer data in a remote data center either as a substitute for, or as a supplement to, customers' on-premises infrastructure*" (Cuni, 2015).

In computer science, the cloud concept is often used to simplify the representation of any network (Mell & Grance, 2011), more or less hiding technical details of the underlying infrastructure because they are either unknown or irrelevant. Also, since cloud applications generally require Internet as a transmission medium, the cloud concept is frequently (mis)interpreted as a kind of synonym of the Internet itself (Strauch et al. 2013). Therefore, in order to clarify the concept of cloud computing, we will shortly describe five main features of cloud computing, four cloud delivery models and three cloud computing service levels.

*2.1 Five Main Characteristics*

Cloud computing describes the benefits of deploying virtualized computing resources over a network, in which the nature and the location of the resource deployment is veiled or not clarified. Resources here mean both computing capacity, data storage, development platforms and software service offerings. These can be adjusted as needed so that only the cost of actual consumption is calculated (Baun et al. 2011). What does this mean? The essential components of cloud computing are characterized in five parts (Figure 1).

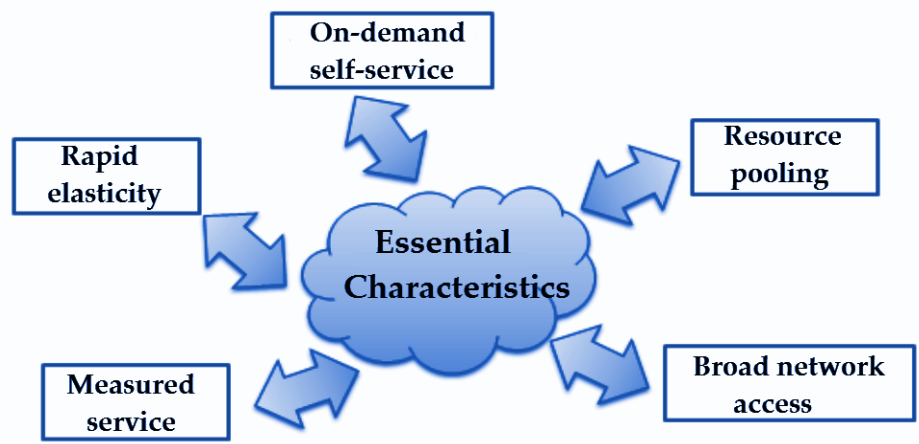

Figure 1. Five features of cloud computing

1. **On-demand self-service:** The first feature is on-demand self-service. This cloud service allows the user to use existing resources, such as Network storage space and server time to automatically adapt to current needs. This dynamic adaptation takes place without any permission request or inclusion of the service employee. The advantage for the cloud provider is that he does not have to hire any or only a small number of employees and still be able to advertise attractive prizes.

2. **Broad network access:** Access to the cloud service must be available or accessible over a network (Internet) connection from the user terminal. The latter can be of various types: mobile devices, tablets, laptops (Leymann, 2009).

3. **Resource pooling:** Resource pooling means that the computing and storage resources of a cloud computing vendor are grouped together to form a resource pool from which different consumers are served with a multi-tenancy model. Different physical and virtual resources are dynamically allocated and adjusted as needed by the cloud consumer. The exact location of the data and resources are unknown to the user.

4. **Rapid elasticity:** Another feature is the elasticity. Upscaling and downscaling can be scaled by intensity of use, avoiding both unused resources and capacity to lose business opportunities. For the cloud user, the available resources of a cloud computing system appear without capacity limit (Leymann, 2009).

5. **Measured service:** The last feature means that the cloud systems automatically measure and control the use of the resource, thus optimizing it if necessary. The measurements of resource use are made publicly available. This leads to transparency and should ensure confidence and security. Depending on the resource, different measurement methods and parameters are used (e.g. memory volume, CPU cycles, bandwidth and number of active user accounts).

*2.2 Four Delivery Models*

Actually, four different cloud delivery models can be distinguished (see Figure 2). The choice of one specific model depends (also) on decisions on privacy and data security criteria.

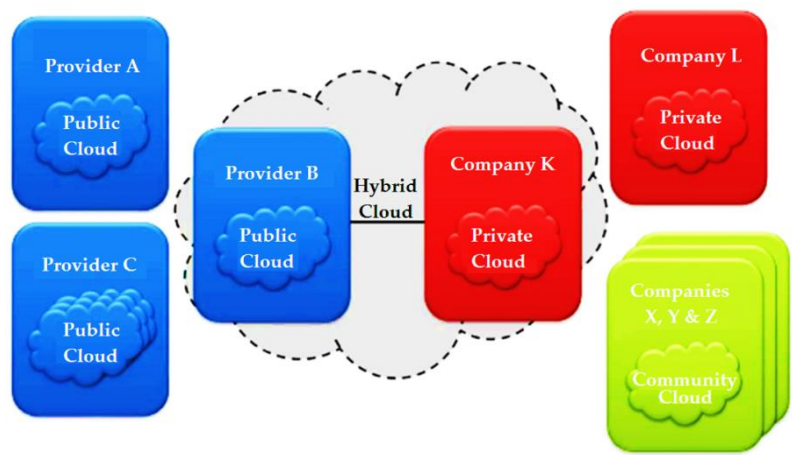

Figure 2. Four delivery models

1. **Private Cloud:** The focus of the private cloud is that "*the provider and the user side belong to the same organizational unit*" (Baun et al. 2011). This model is also referred to in the literature as IntraCloud or InternalCloud. The advantage of this delivery model is that data security issues become obsolete as "*control over the data remains with the user or within the organization*" (Metzger et al. 2011). Only authorized users access the services via intranet or via a virtual private network (VPN).

2. **Public Cloud:** The public cloud is in contrast to the "private cloud" public. An unlimited number of users and operators have access to the services. This success is usually on the internet. The public cloud raises concerns about the security of user data, as they are completely owned by the public cloud provider. Nevertheless, the demand for this offer service is highest and thus has the most users. Payment is made according to the flexible billing model pay-as-you-go, that is, the number of resources consumed.

3. **Community Cloud:** The community cloud is a delivery model that allows the user to access data, e.g. provided by several organizations. This happens when various users of a private cloud service group their data together in a community (Metzger et al. 2011). Here, all members of the community, under their rights, have access to the private cloud of all affiliated organizations.

4. **Hybrid Cloud:** The hybrid cloud is a combination of public and private services. It tries to combine the advantages of the two cloud delivery models and to minimize the disadvantages. Thus, the user has the opportunity to outsource uncritical data in a public cloud, whereas sensitive data with high priority of data security remain in a private cloud (Baun et al. 2011). In doing so, a model has emerged that reacts elastically and dynamically to system load conditions while at the same time denying access to risky data.

*2.3 Three Service Levels*

The architecture of the service models within the cloud is based on the core of the stratification and can be divided into three main levels. Infrastructure as a Service (IaaS) is the lowest level.

Building on this, Platform as a Service (PaaS) follows as the middle layer. The highest level is Software as a Service (SaaS). The layers are arranged according to their degree of abstraction. This should be clarified once again with Figure 3.

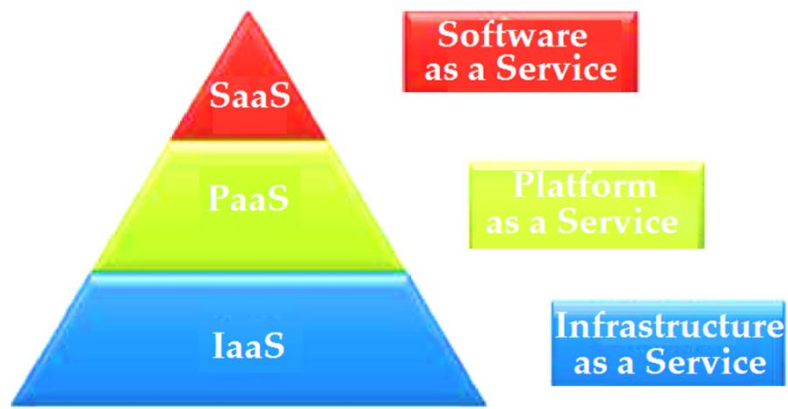

Figure 3. Three cloud service models

1. **Infrastructure as a Service (IaaS):** IaaS is the lowest level in cloud computing and provides the user with an IT infrastructure as a service. This allows him to access virtualized resources such as hard disk capacity (storage), telephone systems, network functions, etc. Furthermore, virtual servers with different operating systems or simply structured databases can be set up by the user in the lowest level. Examples of the IaaS model are Amazon Elastic Compute Cloud (Amazon EC2), Microsoft Windows Azure Platform and HP Cloud Enabling Computing.

2. **Platform as a Service (PaaS):** PaaS is one level higher than IaaS and provides software platforms for application developers but also, simply, to run applications. Developers can access pre-configured programming environments for specific programming languages while executing the developed software in dedicated test environments. Examples of the PaaS model are Microsoft Azure Service, Google App Engine or Force.com.

3. **Software as a Service (SaaS):** SaaS is at the top of the cloud service model. SaaS has been described as "*turn-key operation of application software in the cloud*" (Houssos, 2012). This means, actual applications can be executed by the user via Internet through the web browser. This eliminates local software installations, as well as regular maintenance and updates. Furthermore, there are no additional hardware and license costs for new software releases (Leymann, 2009), and there is minimal or no need for IT expertise by the customer (Houssos, 2012). Typical examples of this model include Microsoft Online Services, Google Apps for Business, Salesforce Customer Relationship Management (Salesforce CRM), Microsoft Live Services, and Google Docs.

*2.4 General Benefits*

The literature and surveys on cloud computing (see for example Gupta et al. 2013; Iyer & Henderson, 2010; Lin & Chen, 2012) describe potential benefits of this new technology for customers, i.e. private customers as well as corporate companies, institutions, service providers, administrations and so on, independent of their sector or size, even if cloud solution may be of particular interest for smaller organizations and institutions: "*Nowadays as the price of storage and bandwidth continues to drop fast, Cloud-Based services are becoming more and more attractive and are affordable to small and medium-sized businesses which are seeking to reduce licensing costs, avoid recruiting IT staff and focus fully on their core responsibility – growing the business*" (Cuni, 2015).

In the following, we will resume nine potential benefits which may be relevant for universities or research organizations interested in CRIS.

1. **(Private) users with inexpensive client computers:** Using cloud computing does not require high performance computers with high acquisition costs. Due to the outsourcing of the data to the cloud and not on local computers, the required computing power is shifted to the servers. Therefore, the client computer is satisfied with simply configured hardware, e.g. a strong processor, but with a small hard drive and low memory. The cost argument includes licensing: a cloud solution means "*less or no initial investment in infrastructure and licences for the possible new costumer*" (Cuni, 2015).

2. **Timeliness of cloud hardware:** SaaS cloud services need to be delivered from the client computer usually only low hardware requirements, since it can be opened and operated with an existing Internet connection, a standard browser and an Internet-enabled device, e.g. notebook, smartphone etc. The hardware of the server computers is constantly adapted by the provider to the rapidly growing technology. It is always up to date, without the user being aware of these extensions.

3. **Timeliness of cloud software:** The software applications are always up to date and automatic. Thus, the user can use the latest updates and features immediately and interactions of the user are spared. Especially for non-IT customers, it's a relief because they do not need to care for complex updates.

4. **Platform independence:** Due to the use of the applications via a conventional web browser, irrespective of which operating system is installed (whether Windows, Linux or Mac OS X), it is possible for the user to access the cloud from any platform. Problems, such as incompatibility between versions, are also insignificant and allow for unrestricted use.

5. **Fast operational readiness:** Cloud applications are usually ready for use immediately after logging in and when they first adapt. Thus, they are convenient and easy to use for the user. They don't need dedicated staff for the maintenance of the system or the infrastructure. In case of questions or problems, support can be offered more agile and fast.

6. **High security against data loss:** The storage of the data is in some cases spread over several locations. This means that the probability of losing data is lower than a backup with usual means, such as data backup on an external hard disk (Krutz & Vines, 2010). Thus, the user does not need to create back-ups which saves additional time. Increased availability and data security are core arguments for the cloud.

7. **Mobile readiness:** In most cases, the cloud application enables use through deployed mobile applications, such as smartphone-based web pages or apps. As a result, the user has access from any location, provided he has an internet-enabled device.

8. **Scalability:** In traditional IT solutions, upscaling is often a costly and time-consuming project. Cloud computing offers an interesting alternative, insofar performant and mature cloud platforms often offer an unlimited scalability by simply pressing a button, whether up or down. Linked to scalability is the fact that a cloud solution "*is paid for use*" (Cuni, 2015), which probably means cost reduction.

9. **Flexibility (customizability):** Most cloud platforms facilitate customized user interfaces. For this, in most cases, no programming skills are needed. Thus, the support of the provider is not confronted with complex change requests and they can deal with more complex adjustments. In other words, "*the client (can) focus their efforts on the core business*" (Cuni, 2015). Or as the Fraunhofer Gesellschaft puts it, "*have corporate responsibility be the driver, not infrastructure governance*" (Küsters & Klages, 2018).

Flexibility is one of the central arguments in favor of cloud computing. "*Applications that run on the cloud take advantage of the flexibility of the computing power available. The computers are set up to work together so that it appears as if the applications were running on one particular machine. This flexibility is a major advantage of cloud computing, allowing the user to use as much or as little of the cloud resources as they want at short notice,* *without any assigning any specific hardware for the job in advance*" (Cuni, 2015). In the following, we'll show what cloud computing means for research information management, based on studies and papers produced by the international community of CRIS providers and customers.

## 3. Cloud as a Challenge for CRIS

A recent global survey conducted by OCLC and euroCRIS on practices and patterns in research information management in several countries revealed that one third of live CRIS are hosted externally, as a cloud solution (Bryant et al. 2018). From 245 responding academic institutions, 88 answered that their research information system is hosted externally (36%). The percentage of cloud solutions depends on the choice of the system provider and the academic environment. In-house developments are most often hosted on campus (89%). In Italy, where the academic infrastructure network Cineca offers an externally hosted system for both research information management and institutional repository (IRIS), most of the IRIS institutions opted for the cloud solution (92%). 37% of the open source DSpace-CRIS solutions are hosted externally. For the most important commercial systems, the survey reveals significant differences: Converis (Clarivate Analytics) is hosted in the cloud for 50% institutions while the percentage for Pure (Elsevier) and Elements (Symplectic) is lower, 39% for Pure and only 13% for Elements.

Some verbatims of the cited survey are interesting; insofar they reflect explicit institutional strategies regarding the cloud. An Australian respondent explained that the choice of a cloud solution was mainly a buying decision, probably based on criteria like human resources (skills), financial considerations (investment) and technology reasons (on campus information technology), and declared that their "*University Policy dictates: 1.) A 'Buy not Build' policy. B.) All applications need to be cloud hosted (…)*". In a quite different context, a Dutch institution highlights that CRIS "*in many institutions and countries are incredibly rich sources of research information (and) represent a huge potential for the international community (…) when linked together into an international research information infrastructure. Not in the least for related research infrastructure initiatives such as the European Open Science Cloud*". Here, obviously, the choice for the cloud has more to do with the integration into an emerging landscape of related information systems and infrastructures. In the following, we'll try to distinguish three different aspects of CRIS and the cloud, the environment of networks and infrastructures, CRIS in the cloud as a service, and the ingestion of cloud data by CRIS.

*3.1 Academic Networks and Infrastructures*

The Dutch verbatim cited above illustrates a situation where an institution is part of a larger network or infrastructure, or wants to join it. This institutional strategy can determine the choice of a cloud solution for the research information management. Yet, on a "lower level", the same strategy can also lead an institution to develop or strengthen the interoperability of its systems.

Especially the European research infrastructure policy has been identified by the CRIS-community nearly ten years ago as an opportunity for the development of research management systems, based on interoperability and standard formats. Keith G. Jeffery from the Rutherford Appleton Laboratory and former president of euroCRIS required in 2012 that CRIS solution should be deployable in virtualized environments like grids and clouds, a development which he considered as a crucial technical future perspective of CRIS (Jeffery, 2012). At the CRIS 2014 conference in Rome, Carl-Christian Buhr (2014) from the European Commission presented the European cloud computing strategy as the "wider digital policy context" for the research and development in the field of research information systems. From 2014 on, the European CRIS network euroCRIS was involved in the EC-funded project HOLA CLOUD, with the purpose of mapping cloud research in Europe (Jeffery, 2014).

Today, the European Open Science Cloud (EOSC[1]) has become one of the eight "pillars of open science" and an essential element of the new Open Science strategy of the European Commission (Brennan, 2018). This has two implications for research information management.

On the one hand, national infrastructures of EU member states are part of the European research cloud. For instance, (Doorn & Dijk, 2012) described the (then) situation of the research infrastructures in the Netherlands and identified one main challenge, i.e. connecting research data to research information and two main barriers, i.e. lacking long-term preservation and lacking interoperability. Also, they argued in favor of international cooperation, especially between the countries of the European research area, to increase standardization and interoperability.

On the other hand, the interconnection of interoperable institutional and national CRIS, based on common standards such as the European CERIF, has the potential to become an "European Research Information Infrastructure", a major provider of metadata on research and thus "*a necessary, complementary, part of the European Open Science Cloud, indispensable to optimally realise the FAIR-aspect (…) Without such a RI infrastructure, providing the needed metadata about the research data in the Cloud, the EOSC will function suboptimally*" (Simons, 2017).

---

[1] EOSC https://ec.europa.eu/research/openscience/index.cfm?pg=open-science-cloud

We already mentioned the Italian academic network Cineca, a not-for-profit consortium with 67 universities, nine research institutions and three university hospitals[1]. As a service provider, Cineca develops information technology applications and systems for the academic sector, including a platform for the research information management called IRIS[2]. Following the results of the OCLC survey, most of the Cineca members use the externally hosted IRIS system, probably not because it was the best choice but because they are already working together as a national consortium and therefore preferred a shared, interoperable solution in the cloud, with a central service provider including further update, helpdesk and maintenance. Moreover, IRIS is compliant with the emerging European "RI infrastructure" insofar the system runs with the CRIS standard format CERIF.

(Clements et al. 2018) describe a kind of partly, networked cloud solution for institutional research information systems in the UK, where the local data are backed-up by a cloud data storage provided by the JISC repository core infrastructure, guaranteeing long-term preservation and access.

Networking as a reason for a cloud solution: our illustration of what this means for research information management put the focus on the European research area. However, networking is not a European invention but a central characteristic of global research, and wherever (and whenever) universities and research organizations work together; they will in some way share information technology and systems. And whenever this cooperation includes research information management, reporting, producing metrics, monitoring etc., cloud solutions will be part of the options.

*3.2 CRIS in the Cloud as a Buying Solution*

We cited above a verbatim from the OCLC survey about "a '*Buy not Build*' policy" which implies that "*all applications need to be cloud hosted*". There are good reasons for research institutions to opt for a cloud solution, and the general benefits of cloud computing mentioned before apply of course also to the special case of CRIS for universities and other research structures. The literature about CRIS provides some examples of CRIS as a cloud solution, from both sides, from system providers as well as from customer institutions.

(Lawson & Herzog, 2014) from UberResearch, a portfolio company from Digital Science, present a shared commercial cloud-based solution for research information management, including a global shared grant database. Other commercial CRIS that are available as a cloud solution are from Elsevier (Pure) and Clarivate Analytics (Converis). Initially, both systems were implemented only as local, on campus applications but they are now also delivered as a cloud solution, as a response to the increasing demand for software and platforms as a service, especially from smaller institutional customers with limited IT expertise.

Based on empirical data on CRIS implementations and projects in German universities and research institutions, a recent survey concludes that in the near future, "*an increasing part of the (research information systems) may move into the cloud, as an infrastructure-, platform- or software-as-a-service solution, similar to other parts of the campus information system infrastructure*" (Azeroual et al. 2019).

(Donohue et al. 2018) from DuraSpace, a not for profit organization that provides leadership and innovation for open technologies, describe another scholarly ecosystem for the management of research information, based on open source software like DSpace (same software as Cineca, see before) and Fedora.

The Spanish SIGMA consortium provides an interesting case study of the development, benefits and requirements of a CRIS solution in the cloud. More than twenty years ago, SIGMA was created by eight Spanish universities as a non-profit software provider for student and research information systems. When SIGMA launched its first fully functional CRIS, it was a kind of "shared in-house development", specifically designed for Spanish universities, and to be implemented as part of the universities' information systems and hosted on the campus, requiring staff, infrastructure, and licenses.

However, over the years, SIGMA had to admit that its solutions "*designed with a very wide functional coverage, (were) suitable for the needs of large universities, but we had a big drawback, anyone who would like to use SIGMA SIS or CRIS solutions should buy the necessary infrastructure and invest on licenses that were not cheap.*

---

[1] Cineca https://www.cineca.it
[2] IRIS https://www.cineca.it/en/content/iris-institutional-research-information-system

*This made difficult for us to enlarge our number of clients*" (Cuni, 2015). For this reason, and "*in order to offer to the markets a more competitive solution*", SIGMA decided to develop a SaaS version of its software, especially for "*relatively small private colleges but of great reputation*".

In a general way, the SIGMA approach illustrates how the development of a cloud solution can be considered as a contribution to strategic marketing with the purpose of gaining new market segments and increasing the number of customers. Regarding the CRIS software, the new approach required "*a review of the organization of the software offered to universities, so as to define lighter suites to be used by higher education colleges and centres, more process-oriented and focused on the end user, who would have also a more specialized and nearer support from SIGMA*" (Cuni, 2015).

Today, SIGMA offers its software in a complete SaaS model for new customers and an upgrade of the implemented software to a cloud solution for the former customers. The success is there: "*Several institutions uses our software as a service*" (Guillaumet, 2015), and today, SIGMA proposes another commercial option for their software, i.e. a consortium model (Guillaumet, 2018). Additionally, SIGMA provides for all customers a private cloud solution whenever asked for.

A virtual private cloud: this is the architecture implemented by the German Fraunhofer Gesellschaft for its institutional CRIS, to cope with their organizational structure of 72 research institutes and units and more than 25,000 staff, and embedded in a complex infrastructure of research information, publications, administrative and research data, library services and so on (Küsters & Klages, 2018).

Following (Houssos, 2012), cloud computing in the field of research information management offers three general benefits:

- "*Minimal or no need for IT expertise by organisations to run the system, very good fit for small-to-medium size institutions (and not only).*
- *Multi-tenant architecture – A single instance serves many clients – Substantial gains in efficiency, utilisation – Facilitates updates / upgrades.*
- *Can be used as a model for running institutional CRIS in a single infrastructure at the regional or national level.*"

On the other hand, the SaaS option of a CRIS requires special efforts, in particular it needs "*considerable effort to develop applications that support multi-tenancy*", i.e. institutions, staff, systems on different locations, with different environments and different demands. Houssos mentions four aspects that need special attention: different identifiers, transparent customization, security (including access control and monitoring/billing), and interoperability with local systems.

Houssos seems relatively confident that implementing a cloud solution can deal with these problems. For Houssos as for the SIGMA team, cloud computing is a serious if not the best option for the development of CRIS. "*100% cloud provides an answer to the challenges that universities and research centers have to face throughout the research lifecycle, from financing to the implementation process and after*" (Guillaumet, 2018).

However, a recent study based on experience at the Czech Technical University in Prague is more skeptical about the ability of cloud solutions to cope with local specific needs and requirements. Here the flexibility of in-house developments is high-lighted: "*An in-house built CRIS has the advantage of a greater flexibility in setting up interfaces with the neighboring systems. It can be more tailored to the needs of the institution, accounting for the organizational or national specificities. This flexibility is limited for all the other options of acquiring a CRIS: contracted bespoke-built systems, CRIS products, or a Software-as-a-Service CRIS offerings. We therefore argue that an in-house built CRIS is better integrated into the information ecosystem of the institution*" (Dvorak et al. 2019).

Perhaps we should add that in fact, (Dvorak et al. 2019) do not question cloud solutions in general but (only) standard (commercial or network) cloud solutions offered and run by external service providers. And second, the Prague experience requires IT skills, investment and staff; and the lack of all these resources is exactly one main reason for the decision in favor of a (external) cloud solution.

*3.3 Data in the Cloud*

The third and last reason to link research information management to the cloud is the fact that more and more data that are required by a CRIS are hosted and available somewhere else on the Internet, "in the cloud". A CRIS must be able to "*import (data) from heterogenous sources to reuse information*" (Guillaumet, 2015). It must cope with virtual research environments (Brasse, 2017) and "knowledge on the web", i.e. the cloud of linked open data (Blümel et al. 2014) or more generally, the "*research content cloud*", including the semantic web, above all Wikipedia via DBpedia (Ghirardi & Scipione, 2014), the "*global digital object cloud*" (Lambert, 2018) or simply the "Big Data" (Rivalle, 2012).

(Brennan, 2018) presents the data model of the CRIS of the Trinity College of Dublin, with different external data sources (e.g. OpenAIRE, DART Europe, BASE, Google Scholar, PeopleFinder, college and school webpages, Scopus, Web of Science, Scival, InCites, PubMed…) and different entry points (institutional repository, library system, web services linked to the CRIS). An integrated CRIS must be able to make use of cloud-based data on research activities, and its strength, so Brennan, is its data model, i.e. the pre-defined interlinkage between several elements, and its accountability guaranteed by the institution.

(Dvorak et al. 2019) provide detailed insight in the "*rich variety of other information systems and services*" with which institutional CRIS "*typically interoperate and interact*", revealing "*several types of heterogeneity: the heterogeneity in the underlying technology, in the data model, the semantic gaps, the different processes*". One part of these information systems is hosted on campus as in-house systems, such as human resources and finance information systems, identity or document management systems, or the institutional open access repository. But other systems are hosted externally, in particular article-level and citation-level bibliographic databases, journal-level citation metrics sources, the global researcher profile registry ORCID, and funder research information systems.

(Dvorak et al. 2019) describe how, at the Czech Technical University in Prague, data are retrieved from these cloud sources and ingested into the local CRIS, through the webservice API the database providers offer, via ad-hoc background jobs set up to contact the external database or ad-hoc imports of MS Excel files. Especially in the case of funder information system, the authors state that they "*are not aware of any established, automated information interchange between institutional and funder CRISs anywhere in the world*". In any case, they conclude that the application of the standard format CERIF would be helpful for the data ingestion from the cloud and that an institutional CRIS could increase the homogeneity and quality of research information retrieved from different, external sources.

Similar to the Czech case study, a survey with 160 German institutions revealed that "*among the external sources, the respondents mention the Web of Science, Scopus, PubMed and the Online Computer Library Center (OCLC) and German National Library catalogues*" (Azeroual et al. 2019).

In the environment of networking institutions, the cloud includes data from other CRIS; in other words, ingesting external data may include the aggregation of data made available by the research information systems from other institutions. Shared standard data models and standard processing of data harvesting and exchange are required (Ghirardi & Scipione, 2014). The same argument is put forward by (Lambert, 2018), underscoring the increasing importance of "*persistent identifiers for a wide range of entities in the research domain*" in the "*global digital object cloud*" and the need to design research information systems in a way that they can deal with multiple identifiers, for instance, by means of a "*federated identifier entity*" added to the data model.

A more formal model of cloud-based data is presented by (Blümel et al. 2014), with seven categories of potential data sources:

- Literature databases,
- search engines,
- individual or institutional websites,
- social networks,
- Wikipedia and Wikidata,
- research information systems,
- research aggregators (like VIVO).

Each category has its own data sources, more or less reliable, and can be characterized with different degrees of internationality, completeness, reusability etc. In particular, institutional and personal websites "*fulfill an important role in academia: the public documentation and communication of any kind of research results*". But because they are "*randomly distributed, heterogeneous, and typically unstructured*", special attention and efforts are required. "*Assuming that in the future highly individualised, heterogeneous web pages of researchers and institutions will be important sources of information about their activities, (the authors) propose to develop solid, easy-to-use software tools for the extraction and provision of structured research information. Such software tools can be used to crawl the web sites of universities and research institutions and to extract metadata from* *them, which is then complemented and consolidated with freely available databases and authority files*" (Blümel et al. 2014).

The ingestion from external data is a challenge for data quality in general and for CRIS quality in particular. The quality of the source data has a direct influence on the quality of the CRIS. (Stempfhuber, 2008) was the first to address data quality as an issue for CRIS; he required that the notion of quality should be made more explicit to support the connection of distributed sources of research information at the level of the European infrastructure cloud. "*All data providing systems have their own, independent data models, describing publications, projects, institutions, people etc., and their curation and maintenance need domain-specific knowledge which must be mobilized for the processing and curation of those data that are exported and ingested into the CRIS*" (Azeroual & Schöpfel, 2019).

Standardization is one major solution to reduce the impact of heterogeneity and inconsistency, with identifiers (ORCID, DOI), standard data formats (CERIF) and terminology (CASRAI). Typical data errors and new techniques and methods of data cleansing and monitoring in order to control and increase the quality of ingested data have been summarized and discussed by (Azeroual & Schöpfel, 2019).

## 4. Implementing CRIS in the Cloud

*4.1 Service Models*

Universities and research institutions are increasingly interested in CRIS services in the cloud. However, the choice of the specific kind of service depends on the local needs and requirements.

Based on the fact that software and IT infrastructures are operated by an external IT service provider and the user uses CRIS as a service, the combination of both technologies can also be described as CRIS as a Service (CRISaaS). The architecture of CRIS consists of three different levels, which are differentiated as follows (Azeroual & Schöpfel, 2019):

- **data collection** (internal and external data sources from different systems are collected),
- **data integration** / **data storage** (the integration of research information is filled and stored in the CRIS by means of an ETL process (extraction, transformation, load)) and
- **data presentation** (research information is analyzed in CRIS and provided to the end-user).

Above, we have described three different cloud service levels. Also, we can distinguish three different cloud service models for CRIS.

**Infrastructure as a service**. The scientific institutions have already outsourced part of their data sources to an external IT service provider, so the CRIS solution for data collection usually offers vendor-specific interfaces from the infrastructure level. The CRIS process steps of data integration and data storage can be considered platforms for CRIS. The CRIS platforms are then used in the data presentation, the application layer, to display the evaluated data (Figure 4).

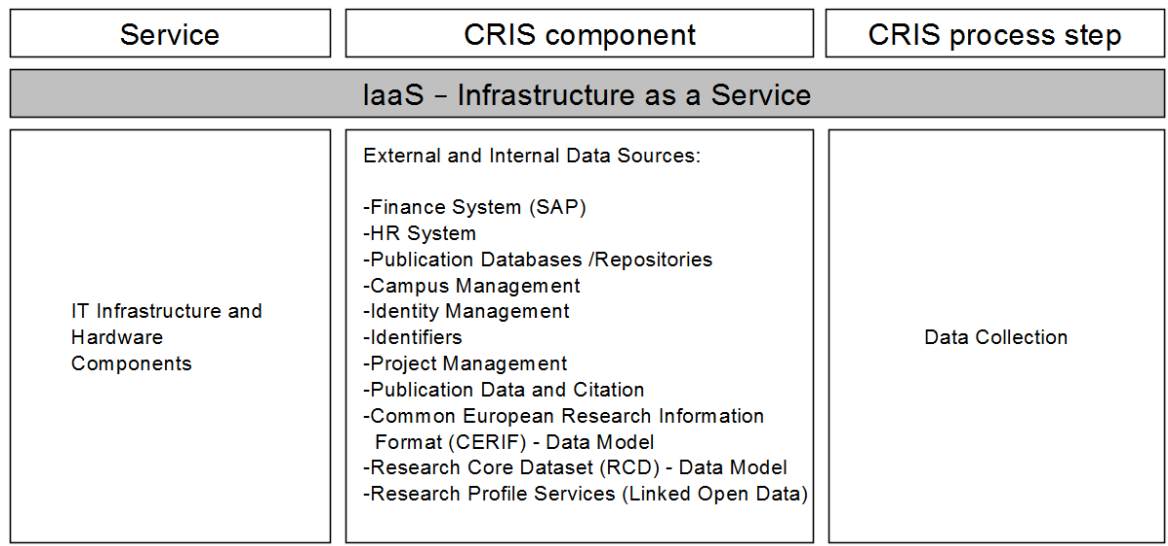

Figure 4. CRIS in IaaS level

**Platform as a service**. Within the PaaS level, integrating source data into a multi-dimensional data model and executing complex queries place high demands on the performance of the IT component involved (Seufert & Bernhardt, 2010). Because such queries usually not continuously but sporadically take place, to evaluate research information of an institution. This shows the strengths of the CRISaaS solution.

The scalability of a cloud infrastructure allows organizations to quickly leverage additional analytical capacity for analytical queries without having to permanently operate a reserve capacity-based CRIS infrastructure (Figure 5).

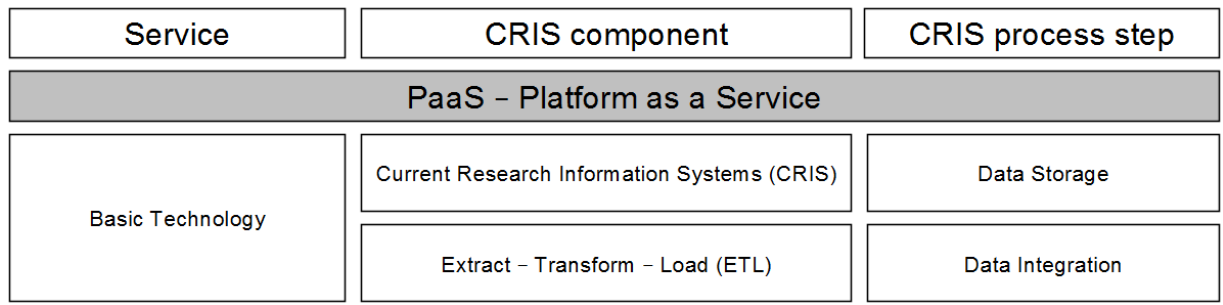

Figure 5. CRIS in PaaS level

**Software as a service**. At the SaaS level, this means that CRIS application components can be conveniently and quickly expanded to the desired functionality. Thus, the scope of applications can be flexibly adapted to all scientific institutions.

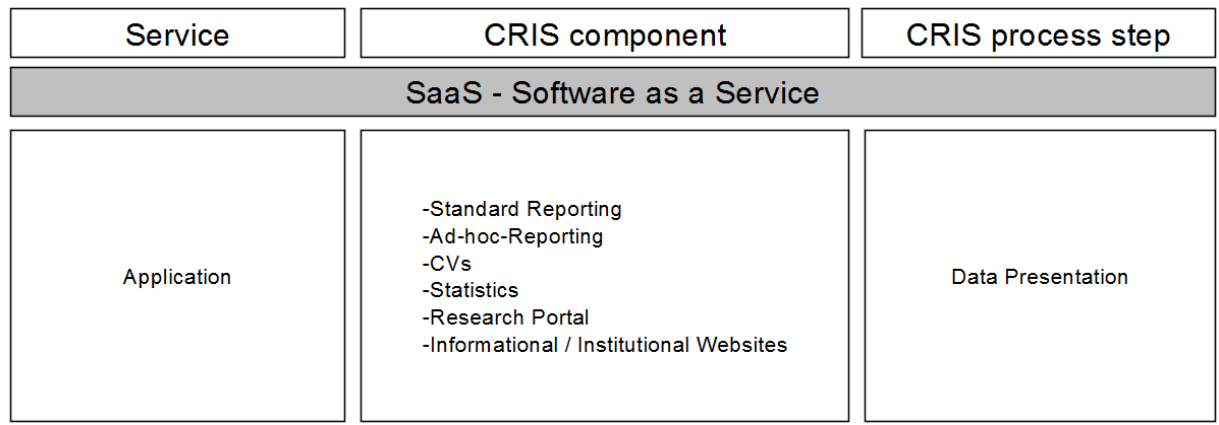

Figure 6. CRIS in SaaS level

In the exemplary Public CRIS cloud architecture, the user's operating source systems are also operated in a cloud solution. The processed data is accessed via a web browser. All CRIS services and CRIS components are provided by the CRIS cloud provider. These include Extraction Transformation Load (ETL) tools, CRIS as well as evaluation tools of business intelligence such as (reporting).

Depending on the requirements of the facilities, the operating modes of the private or hybrid CRIS cloud are also conceivable, in which the source systems are completely or partially at the customer and only one analyzer-relevant data section is transmitted to the CRIS provider.

*4.2 Requirements*

Figure 7 illustrates a complete CRIS solution in the cloud.

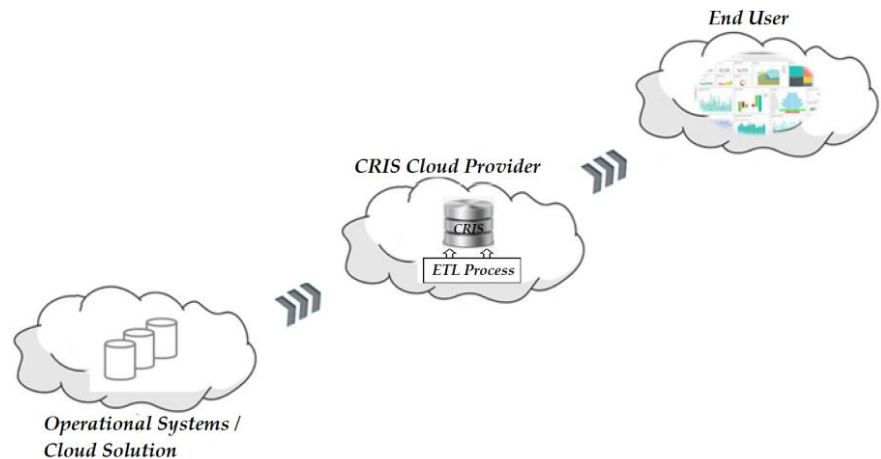

Figure 7. CRIS Cloud Architecture

Before institutions decide on a CRIS, various aspects need to be considered first. For this reason, in this section, the requirements, in particular for cloud-based CRIS are shown.

**Introduction time:** Universities and research institutions have limited financial and human resources. As a result, limited resources are available for IT projects. In this sense, a rapid deployment of the CRIS solution becomes a priority request to save valuable resources and quickly take advantage of the solution found.

**Provision of information:** Depending on the requirements of the user, information must be provided in different contexts. As a rule, this takes the form of reports and analyzes with further options for formatting. For easy sharing of documents, their results should be exportable in Microsoft Office formats.

**Usability:** CRIS must offer a different range of functions in terms of the tasks to be performed and the addressed user group. This ranges from viewing static reports for occasional users. Despite the proposed functions, the user tools must be intuitive and without time-consuming user training to operate.

**Data quality:** Irrespective of the chosen delivery model of a CRIS solution, data quality is an essential prerequisite for generating decision-relevant information from operational data. Inadequate data quality inevitably leads to user loss of trust. That's why everyone involved in the operational systems must produce high quality data.

**Processing speed:** The processing speed is influenced by the query and load performance of the CRIS. Thus, complex user queries but also frequent data extraction from the source systems into the CRIS can lead to a reduction in system performance. To keep system performance up to the user's requirements, operating a CRIS solution in the cloud requires a constant Internet connection with sufficient bandwidth for data transmission.

**Compatibility:** The integration of heterogeneous source systems is a comprehensive data analysis in an institution of great importance. Consequently, for the transfer of the data into the CRIS, it must be checked whether the CRIS platform offers various interfaces or import functions for the smooth exchange of data between the application systems.

*4.3 Risks and Barriers*

Unfortunately, despite the many benefits that cloud computing brings to CRIS users, there are also disadvantages to using cloud services. We already mentioned critics from (Dvorak et al. 2019) concerning limited flexibility and autonomy. Other reasons against using cloud computing are:

**Security:** The topic of security is a controversial topic; on the one hand users think that it is not safe to leave important data in hand of the providers, because the probability of data theft, data manipulation or data loss is high. On the other hand, data centers of cloud providers often offer a higher level of security. Data security should also be handled with great care and attention. Because security concepts must implement measures that ensure the data at a high and reliable level. The level of security is variable per provider, so it is important to make a proper choice of the provider.

**Finding suitable providers and dependency of the provider:** It is very important to find the right provider. A suitable supplier is reliable, and must have a good reputation, so it is easy to have a predictable and stable cooperation. For the smooth operation of the cloud services, the determination of technical conditions and the integration of the new processes with existing processes are essential. This is one of the selection criteria of a provider.

A key aspect of choosing cloud providers that should always be respected is the creation of dependencies. It should be avoided that a subsequent change of cloud provider because of dependencies with the current provider would no longer be possible. The more independent an application is from the cloud provider, the better. It should also be noted that future integration with other applications must be possible so that the growth of the product (project, application, service, etc.) is not affected.

It should be noted that there are still some standards missing for cloud platforms, so there are difficulties in changing providers, which can sometimes be costly and time consuming. One solution would be to think about and work out about exit strategies.

Of course, if there are different problems in providing the service, the provider is responsible for it, but the user has to be seen by the public in general. Sometimes it can mean that something has to be paid in addition.

**Legal questions:** There are still many legal issues that remain inadequately resolved, as well as lack of performance and contract standards. These legal issues are a barrier to cloud computing. Notwithstanding, there are established cloud providers offering services in line with their contract. Compliance with legal regulations is a decision criterion for choosing a cloud provider. In Europe, for example, data protection laws apply.

Unless there are widely recognized international certificates for cloud service providers, the following is important to the cloud user: standards and certifications for cloud service providers should be consistent with existing approaches that are generally accepted in the industry, such as: based on ISO 27001 and ISO 27002, which are supplemented by essential cloud specifics.

**Confidence:** In order to solve the problem of trust, it is important to conclude individual contracts. The contract contains the terms and conditions, the service description, the points of flexibility and scalability. For this, the service level agreements are necessary, there is exactly specified which services are to be provided and which sanctions apply, in the case of non-performance.

**Unverifiable data storage:** The data management is a responsibility of the provider, so it is in principle not verifiable by the customer. It is not controllable, for example, whether the data is deleted successfully, if the customer has used a corresponding command, or if the redundant copies of the data are deleted.

**Data protection:** The development of cloud computing, like many other technologies, poses unanswered questions about the management of personal information. The big problem comes from the nature of the Internet, because the Internet is a worldwide network of computer networks that operate autonomously. Each computer can be located in another country. Therefore, not only the laws of one country (for example, Germany) but also the laws of all involved in the transmission and processing of data must be considered.

## 5. Conclusion and Outlook

In less than ten years, cloud computing has become an important and significant aspect of research information management, as part of academic networking and research infrastructures, as provider of essential data for research information systems, and as a solution for these systems themselves. Our paper shows that the choice of a cloud solution for a CRIS project is not only a technical question but must be considered in a complex environment or ecosystem including different systems and tools, stakeholders and users, and legal and security issues. This global approach must constantly address the problem of the security constraint that must adapt to the dynamics of the environment, of changing needs, risks, and technologies.

The finality of such an approach will make it possible to sustain the CRIS platform via a number of key parameters while ensuring operational flexibility. Some aspects:

- CRIS researchers and administrators will manage only the application layer of the research platform through the extraction and exploitation of data and results, while the cloud architecture provider will ensure the operation and maintenance of the research platform services.

- The CRIS database can be segmented and deployed on multiple servers, each server responding to a specific need.

- The use of cloud access procedures will not only contribute to a better management of human workload but will undeniably provide better traceability and consolidate the security protocol specific to these types of systems.

- This approach can be adapted and deployed on different research platforms in particular to control access to sensitive information.

On the other hand, cloud computing technology offers many benefits to users of research information systems, such as increased data security, increased organizational flexibility, lower IT administration costs and reduced costs.

To sum up, the potential of analytical applications in the cloud is great. Especially the cost savings and performance improvements as well as the new dimension of data integration make the field of the CRIS cloud very interesting and offer new opportunities for many scientific institutions.

For universities and research institutions that can no longer satisfy their analytical needs exclusively through the use of spreadsheets and inhouse databases, CRIS cloud solutions provide fast and cost-effective access to advanced functionalities, ranging from basic automated reporting capabilities to the distribution of reports and analytics and to the integration of data from multiple sources, together with increased control of data quality.

The targeted combination of the two technologies CRIS and cloud eliminates the cost of implementing and maintaining a CRIS solution in the facility's own infrastructure. It also provides further savings by eliminating the need for local software installations, including the necessary licensing.

With regard to the feasibility of a CRISaaS solution for universities and research institutions, it should be noted that institutions that already use a large proportion of their source systems as SaaS solution benefit from a corresponding CRIS solution. Furthermore, merging and analyzing data from hosted legacy systems within a CRISaaS solution does not pose an additional data security risk. Thus, cloud computing offers an alternative to a traditional on-premise CRIS solutions; a CRISaaS solution, under the conditions mentioned, appears as a feasible option for small, medium or large institutions.

It is expected that the dissemination of CRIS cloud solutions to universities and research institutions will increase in the next few years and thus more references from users will be available. Due to the trend to outsource part or all of IT and provide and use it as a service, more and more services will be available in the cloud. Accordingly, formerly on-premise data sources move into the cloud, which can be integrated via standardized interfaces within a CRISaaS solution.

This development is supported by a dynamic market of different categories of CRIS cloud providers, which makes CRIS technology available to a broader target group. Therefore, universities, research institutes and scientists have to deal with the subject in good time and face the technical and safety-critical challenges in order not to lose touch with their competitors.

Following our overview and in order to accompany the future development of research information management in the cloud, a couple of topics need further attention and assessment, especially in the emerging ecosystem of open science, such as:

- How do CRIS cloud solution deal with the concepts of transparency and integrity of the research and evaluation process?
- How do they cope with changing requirements of national and international laws (privacy etc.) and with new conditions of funding agencies, research organizations and authorities?
- How do they take into account the emerging research infrastructures such as the European Open Science Cloud (EOSC)?
- How do they interconnect with research data repositories or other research data services; what is the impact of the FAIR-principles of research data management (Wilkinson et al. 2016) on research information management?

Also, more case studies of the preparation, implementation and management of cloud-based CRIS may be helpful to provide further empirical evidence for key factors of success, for risks and barriers but moreover, for good practice in universities and research institutions.